# Regression prediction algorithm for energy consumption regression in cloud computing based on horned lizard algorithm optimised convolutional neural network-bidirectional gated recurrent unit

Feiyang Li[1], Zinan Cao[2], Qixuan Yu[3], Xirui Tang[4*],
[1]Department of Computer Science, University of Illinois Urbana-Champaign, Champaign, IL, 61820, USA
[2]Department of General Systems Studies, The University of Tokyo, Tokyo, 113-8654, Japan
[3]College of Computing, Georgia Institute of Technology, Atlanta, GA, 30332-0280, USA
[4]College of Computer Sciences, Northeastern University, Boston, MA, 02115, USA
* Corresponding author: e-mail: xirtang@gmail.com

*Abstract*—For this paper, a prediction study of cloud computing energy consumption was conducted by optimising the data regression algorithm based on the horned lizard optimisation algorithm for Convolutional Neural Networks-Bi-Directional Gated Recurrent Units. Firstly, through Spearman correlation analysis of CPU, usage, memory usage, network traffic, power consumption, number of instructions executed, execution time and energy efficiency, we found that power consumption has the highest degree of positive correlation with energy efficiency, while CPU usage has the highest degree of negative correlation with energy efficiency. In our experiments, we introduced a random forest model and an optimisation model based on the horned lizard optimisation algorithm for testing, and the results show that the optimisation algorithm has better prediction results compared to the random forest model. Specifically, the mean square error (MSE) of the optimisation algorithm is 0.01 smaller than that of the random forest model, and the mean absolute error (MAE) is 0.01 smaller than that of the random forest.3 The results of the combined metrics show that the optimisation algorithm performs more accurately and reliably in predicting energy efficiency. This research result provides new ideas and methods to improve the energy efficiency of cloud computing systems. This research not only expands the scope of application in the field of cloud computing, but also provides a strong support for improving the energy use efficiency of the system.

*Keywords-Random Forests; Cloud Computing; Horned Lizard Algorithm Optimisation;*

## I. INTRODUCTION

Cloud computing, as an elastic and efficient computing model, has been widely used in various fields. However, as the scale of cloud computing continues to expand and the number of data centres increases, energy consumption becomes an increasingly serious problem [1]. Energy consumption in data centres not only negatively affects the environment, but also increases operational costs, so how to improve energy efficiency in cloud computing environments has become one of the hotspots in current research.

When studying energy efficiency in cloud computing, we usually focus on parameters such as CPU usage, memory usage, network traffic, power consumption, number of executed instructions and execution time [2]. There are complex correlations and influences between these parameters, and traditional optimisation methods often find it difficult to comprehensively consider the interactions between these factors. In contrast, machine learning, as a powerful data analysis tool, can help optimise energy efficiency in cloud environments by learning and modelling massive data and discovering hidden patterns and trends [3].

Machine learning algorithms play an important role in energy efficiency research. Firstly, by analysing and mining historical data, predictive models can be built to forecast future system resource utilisation, which in turn can adjust the system configuration to improve energy utilisation [4]. Second, machine learning can also optimise resource allocation through intelligent scheduling algorithms to reduce power consumption while ensuring system performance. In addition, machine learning can also monitor the system operation status in real time through the intelligent decision-making system and adjust the system parameters according to the real-time data to achieve the best energy efficiency.

In summary, the study of energy efficiency in cloud computing is one of the urgent problems that need to be solved nowadays. Machine learning algorithms, as a powerful tool, play an increasingly important role in this field, and are expected to help us better optimise resource utilisation in cloud environments, reduce energy consumption, and promote the application and dissemination of the concept of sustainable development in cloud computing. In this paper, we provide a new way of thinking for predicting energy consumption in cloud computing based on the data regression algorithm of Horned Lizard Optimisation Algorithm for Optimising Convolutional Neural Networks - Bi-directional Gated Recurrent Units.

## II. RELATED WORK

The dataset used in this paper is selected from the open source dataset, which includes parameters such as CPU usage, memory usage, network traffic, power consumption, number of instructions executed and execution time, with energy efficiency as the target variable. It can be used to explore the impact of machine learning optimisation techniques on energy efficiency in cloud environments. Selected part of the data is shown in Table 1.

TABLE I.  SELECTED DATA SETS

| Cpu usage | Memory usage | Network traffic | Power consumption | Execution time | Energy efficiency |
|---|---|---|---|---|---|
| 54.88 | 78.95 | 164.78 | 287.81 | 69.35 | 0.55 |
| 43.76 | 22.46 | 429.14 | 272.96 | 60.15 | 0.46 |
| 38.34 | 16.44 | 779.79 | 382.76 | 42.16 | 0.14 |
| 79.17 | 2.97 | 926.37 | 173.56 | 55.70 | 0.78 |
| 56.80 | 2.36 | 722.55 | 143.34 | 79.70 | 0.94 |
| 7.10 | 96.52 | 919.17 | 275.63 | 39.97 | 0.85 |
| 2.02 | 89.34 | 208.42 | 199.26 | 61.85 | 0.70 |
| 11.83 | 17.49 | 433.68 | 214.91 | 12.59 | 0.11 |
| 41.47 | 74.77 | 757.37 | 96.01 | 77.19 | 0.42 |
| 61.69 | 0.67 | 686.37 | 154.89 | 99.54 | 0.99 |
| 35.95 | 72.98 | 350.51 | 132.66 | 76.33 | 0.00 |
| 6.02 | 33.04 | 719.80 | 434.59 | 28.99 | 0.79 |
| 67.06 | 56.20 | 856.37 | 300.82 | 1.63 | 0.98 |
| 36.37 | 21.96 | 765.11 | 300.07 | 44.71 | 0.69 |

## III. SPEARMAN CORRELATION ANALYSIS

Spearman's correlation analysis is a non-parametric statistical method used to measure the monotonic relationship between two variables. It is used to measure the correlation between the rank order of two variables without requiring the relationship between the variables to be linear. Spearman correlation analysis is performed on CPU, usage, memory usage, network traffic, power consumption, number of instructions executed, execution time and energy efficiency, the correlation heat map is shown in Fig. 1 and the correlation rank order of each parameter with energy efficiency is output and the result is shown in Fig. 2.

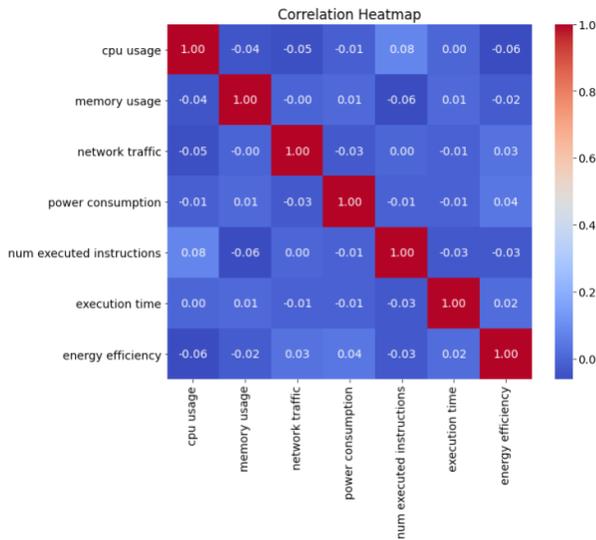

Figure 1. Correlation heat map.

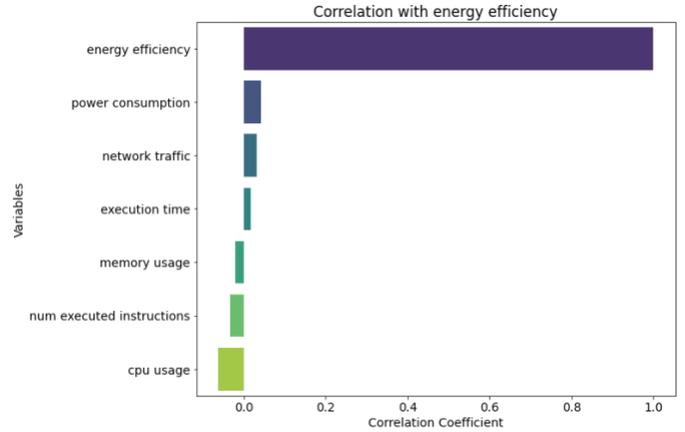

Figure 2. The correlation rank order of each parameter with energy efficiency.

From the correlation heat map, it can be seen that the largest positive correlation with energy efficiency is power consumption, and the highest negative correlation with energy efficiency is CPU utilisation.

## IV. METHOD

### A. Random forest

Random Forest is an integrated learning method that achieves tasks such as classification and regression by constructing multiple decision trees. Its principle is based on the idea of Bagging and random feature selection [5].

Firstly, Random Forest generates multiple decision trees by Bagging method, which is a self-sampling technique where a number of samples are taken from the original dataset in a putative way and used to train each decision tree. This results in multiple decision trees that are independent of each other and reduces the risk of overfitting.

Second, in the training process of each decision tree, Random Forest introduces the mechanism of random feature selection. In each node division, only a random subset of features is considered for division, instead of all features [6]. This randomness makes the individual decision trees more different from each other and improves the generalisation ability of the overall model.

Finally, in the prediction phase, for the classification task, the random forest determines the final classification result by voting (majority vote); for the regression task, the average of each decision tree is taken as the final prediction result. Since each tree is trained based on part of the data and features, the overall model has good generalisation ability and overfitting resistance.

### B. Optimisation algorithm

The data regression algorithm based on the Horned Lizard Optimisation Algorithm for Optimising Convolutional Neural Networks - Bidirectional Gated Recurrent Units (CNN-BiGRU) is an innovative approach that combines a bio-heuristic optimisation algorithm and a deep learning model. In this algorithm, HLOA is used to optimise the parameters of the CNN-BiGRU model to improve the performance of the data regression task [7].

CNN is a deep learning model specifically designed to process image data by extracting features of the image through convolutional operations [8]. BiGRU, on the other hand, is a recurrent neural network structure for sequence data processing that captures long-range dependencies in sequence data. Combining these two models allows for better processing of data that contains both spatial and temporal information [9].

The schematic diagram of the Horned Lizard Optimisation Algorithm is shown in Figure 3. And HLOA, as a bio-inspired optimisation algorithm, is inspired by the ability of horned lizards to react and adapt quickly to environmental changes in nature.HLOA has a hierarchical structure and multiple optimisation strategies, which is able to balance the global exploration and local exploitation during the search process, so as to find the optimal solution efficiently [10].

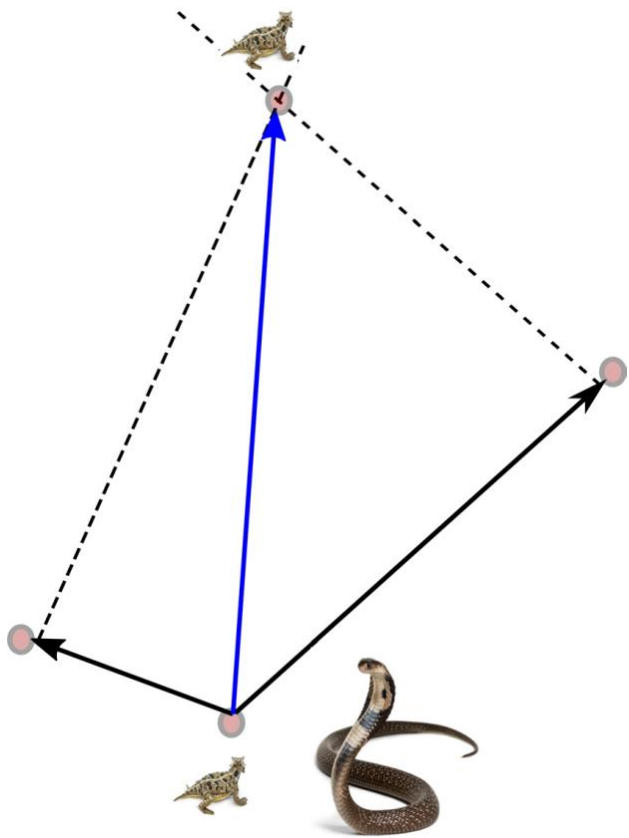

Figure 3. The schematic diagram of the sparrow search algorithm.

In applying HLOA to optimise the CNN-BiGRU model, the fitness function, which is a measure of model performance, is first defined. The mean square error is chosen as the fitness function. Then, the hyperparameters and weight values in the CNN-BiGRU model are continuously adjusted using the HLOA algorithm so that the fitness function reaches the minimum value or converges to a satisfactory level.

Through the continuous iterative optimisation process, HLOA can help the CNN-BiGRU model to better fit the training data, generalise to the test data, and achieve better prediction results in data regression tasks. This approach, which combines bio-heuristic algorithms and deep learning models, has potential applications in dealing with complex data regression problems and shows strong performance performance and robustness.

V. METHOD

This experiment was conducted using matlab to test the data regression algorithm models introducing random forest model and horned lizard based optimisation algorithm for optimising convolutional neural network-bi-directional gated recurrent unit respectively.Parameters such as MSE, RMSE, MAE, MAPE and $R^2$ were used to evaluate the model effectiveness.

MSE is the mean of the squares of the differences between the predicted and true values, the smaller the MSE the better the model fit, RMSE is the square root of MSE, RMSE is the same as the original data in terms of magnitude, which gives a more intuitive understanding of the magnitude of the prediction error, MAE is the mean of the absolute values of the differences between the predicted and true values, MAE measures the extent to which the predicted values deviate from the true values, the smaller the better, MAPE is the mean of the relative values between the predicted and true values, MAPE is the mean of the relative values between the predicted and true values, MAPE is the mean of the relative values between the predicted and true values. is the mean percentage of the absolute value of the relative error between the predicted value and the true value.MAPE can help evaluate the predictive accuracy of a model on different scales.$R^2$ represents the proportion of the dependent variable (target variable) that can be explained by the independent variable.$R^2$ ranges from 0 to 1, with the closer to 1 the better the model is fitted.

The results of the evaluation parameters for the random forest and optimisation algorithms are shown in Table 2 and Figure 4.

TABLE II. THE RESULTS OF THE EVALUATION PARAMETERS

| Model | MSE | RMSE | MAE | MAPE | $R^2$ |
|---|---|---|---|---|---|
| Random forest | 0.119 | 0.41 | 0.289 | 59.797 | -0.093 |
| Optimisation algorithms | 0.10925 | 0.33053 | 0.27646 | 2.9941 | -0.27001 |

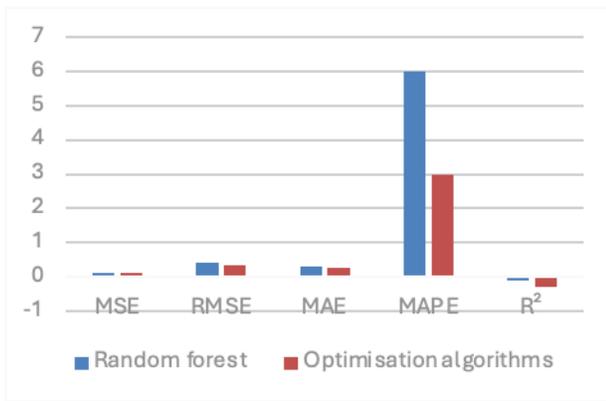

Figure 4. The results of the evaluation parameters.

As can be seen from the model evaluation indicators, the MSE of the optimisation algorithm is smaller than the random forest model by 0.01, and the MAE of the optimisation algorithm is smaller than the random forest by 0.013, and the combined results of all the indicators show that the optimisation algorithm has a better prediction than the traditional random forest algorithm, and is able to more accurately predict the energy efficiency.

Output the optimisation algorithm test set prediction value and the actual value of the line graph, the results are shown in Figure 5.

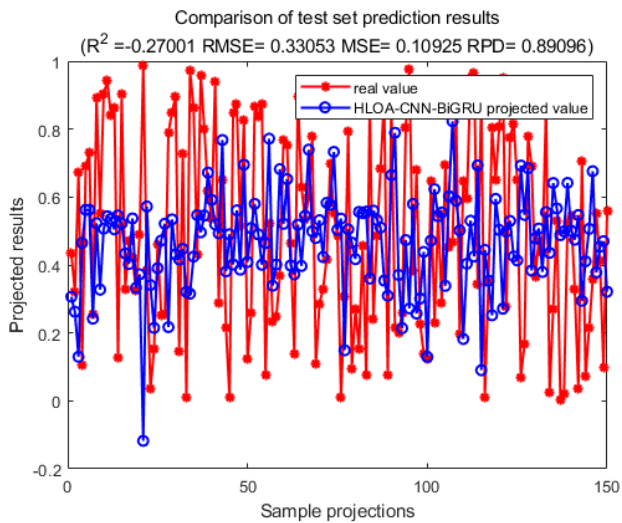

Figure 5. Line graph of projected-actual values.

This paper optimises the data regression algorithm of convolutional neural network-bidirectional gated recurrent unit based on the horned lizard optimisation algorithm for predicting energy consumption in cloud computing, which provides a new idea for studying energy efficiency in the field of cloud computing. Firstly, through Spearman correlation analysis of CPU, usage, memory usage, network traffic, power consumption, number of executed instructions, execution time and energy efficiency, we found that power consumption has the highest degree of positive correlation with energy efficiency, while CPU usage has the highest degree of negative correlation with energy efficiency. In our experiments, we tested the random forest model and the convolutional neural network-bidirectional gated recurrent unit data regression algorithm model optimised based on the horned lizard optimisation algorithm using Matlab. According to the model evaluation metrics, the optimisation algorithm reduces the mean square error (MSE) by 0.01 and the mean absolute error (MAE) by 0.013 compared to the random forest model. The results of the combined metrics show that the optimisation algorithm is significantly better than the traditional random forest algorithm in predicting the energy efficiency more accurately.

In other words, through the optimisation algorithm proposed in this study, significant results have been achieved in the prediction of energy consumption in the field of cloud computing. This study not only deeply explores the correlation between the indicators, but also successfully improves the traditional model by introducing the horned lizard optimisation algorithm, which improves both prediction precision and accuracy. Therefore, this study provides an effective way and method to improve the energy utilisation efficiency of cloud computing system, which is of positive significance for promoting the development of cloud computing field. It is hoped that the sample size can be further expanded and the applications in different scenarios can be studied in depth in the future, with a view to better serving the needs of actual production and life and promoting cloud computing technology to play a greater role in sustainable development.